\documentclass[12pt,thmsa]{article}
\usepackage{amssymb}
\usepackage{graphicx}

\input{tcilatex}

\begin{document}

\begin{center}
{\Large Transient spectrum of a single-Cooper-pair box with binomial states }

\bigskip

\textbf{Mahmoud Abdel-Aty$^1$\footnote[3]{%
Corresponding author: \textbf{abdelatyquantum@yahoo.co.uk}}, H. F.
Abdel-Hameed$^2$ and N. Metwally$^2$}

$^1$Mathematics Department, College of Science, Bahrain University,
Kingdom of Bahrain

 $^2$Mathematics Department, Faculty of Science,
South Valley University, Egypt
\end{center}

We present an analytical expression for the response of a transient spectrum
to a single-Cooper-pair box biased by a classical voltage and irradiated by
a single-mode quantized field. The exact solution of the model is obtained,
by means of which we analyze the analytic form of the fluorescence spectrum
using the transitions among the dressed states of the system. An interesting
relation between the fluorescence spectrum and the dynamical evolution is
found when the initial field states are prepared in binomial states.

\section{Introduction}

For quantum information science and technologies, it is crucial to build the
fundamental quantum logic gates \cite{str03}. Together with the basic single
bit logic gates, the non-trivial two bit gates constitutes the fundamental
blocks for the quantum network of quantum computing. The present lack of a
current standard based on quantum devices has inspired several attempts to
manipulate single electrons, or Cooper pairs, where the rate of particle
transfer is controlled by an external frequency. Various superconducting
nanocircuits have been proposed as quantum bits (qubits) for a quantum
computer \cite{ber03,wei04}. In principle, any two-state quantum system
works as a qubit, the fundamental unit of quantum information. However, only
a few real physical systems have worked as qubits, because of requirements
of a long coherent time and operability. Among various physical
realizations, such as ions traps, QED cavities, quantum dots and NMR etc.,
superconductors with Josephson junctions offer one of the most promising
platforms for realizing quantum computation \cite%
{ber03,wei04,pla03,ave03,wal04}.

The single-Cooper-pair transistor \cite{ful89,yu01}, is composed of two
ultrasmall tunnel junctions in series forming an island. Transport through
the single-Cooper-pair transistor depends on the electrostatic energy
required to charge the island, as in the single-electron transistor, and
also on the Josephson coupling across the junctions. Cooper-pair boxes are
one of the prominent candidates for qubits in a quantum computer. Recent
experiments \cite{pas03} have revealed quantum coherent oscillations in two
CPBs coupled capacitively and demonstrated the feasibility of a conditional
gate as well as creating macroscopic entangled states. Scalable
quantum-computing schemes \cite{pas03} have been proposed based on charge
qubits. In architectures based on Josephson junctions coupled to resonators,
the resonators store single qubit states, transfer states from one Josephson
junction to another, entangle two or more Josephson junctions, and mediate
two-qubit quantum logic. In effect, the resonators are the quantum
computational analog of the classical memory and bus elements.

In this paper we deal with the problem of the interaction between a
single-mode quantized field and a single-Cooper-pair box biased by a
classical voltage. Despite the complexity of the problem, we obtain a quite
simple master equation that is valid for arbitrary values of the Rabi
frequency and the detuning. We apply the Fourier transform of the time
averaged dipole-dipole correlation function to calculate the fluorescence
spectrum, assuming that the electromagnetic field is initially in a binomial
state. We find that the detuning changes considerably the shape of the
resonance fluorescence spectrum and leads to novel spectral features. The
organization of this paper is as follows: in section 2 we introduce the
model and give exact expression for the unitary operator. In section 3 we
employ the analytical results obtained in section 2 and by using the finite
double-Fourier transform of the two-time field correlation function we find
an analytical expression for the spectrum. Finally, we summarize the results
in section 4.

\section{The model}

Several schemes have been proposed for implementing quantum computer
hardware in solid state quantum electronics. These schemes use electric
charge, magnetic flux, superconducting phase, electron spin, or nuclear spin
as the information bearing degree of freedom \cite{mak99}. In this paper, we
consider an example of a realistic system, fabricated by the present day
technology. We consider a superconducting box with a low-capacitance
Josephson junction (with the capacitance $C_{J}$ and Josephson energy $E_{J}$%
), biased by a classical voltage source $V_{g}$ through a gate capacitance $%
C_{g}$ and placed inside a single-mode microwave cavity. Suppose the gate
capacitance is screened from the quantized radiation field, then the
junction-field Hamiltonian, in the interaction picture, can be written as
\cite{mig01}
\begin{equation}
\hat{H}_{in}=\frac{(Q-C_{g}V_{g}-C_{J}V)^{2}}{2(C_{g}+C_{J})}-E_{J}\cos \phi
.  \label{5}
\end{equation}%
where the relevant conjugate variables are the charge $Q=2Ne$ on the island
(where $N$ \ is the number of Cooper-pairs) and the phase difference $\phi $
across the junction. The radiation field is to produce an alternating
electric field of the same frequency across the junction, and $V$ is the
effective voltage difference produced by the microwave across the junction.
We assume that the dimension of the device is much smaller than the
wavelength of the applied quantized microwave (which is a realistic
assumption), so the spatial variation in the electric field is negligible.
We also assume that the field is linearly polarized, and is taken
perpendicular to the plane of electrodes, then $V$ may be written down as
\cite{zha02}
\begin{equation}
V=i\left( \frac{\hslash \omega }{2C_{F}}\right) \left( \hat{\psi}-\hat{\psi}%
^{\dagger }\right) ,
\end{equation}%
where $\hat{\psi}^{\dagger }$ and $\hat{\psi}$ are the creation and
annihilation operators of the microwave with frequency $\omega $. $C_{F}$ is
the capacitance parameter, which depends on the thickness of the junction,
the relative dielectric constant of the thin insulating barrier, and the
dimension of the cavity. In this paper we consider the case where the
charging energy with scale
\[
E_{c}=\frac{e^{2}}{2}\left( C_{g}+C_{J}\right) ,
\]%
dominates over the Josephson coupling energy $E_{_{J}}$, and concentrate on
the value $Vg=\frac{e}{C_{g}}$, so that only the low-energy charge states $%
N=0$ and $N=1$ are relevant. In this case the Hamiltonian in a basis of the
charge state $\left\vert \downarrow \right\rangle $ and $\left\vert \uparrow
\right\rangle $ reduces to a two-state form. In a spin-$1/2$ language \cite%
{kre00}
\begin{equation}
\hat{H}_{in}=E_{c}\left( 1+\frac{C_{J}^{2}}{e^{2}}V^{2}\right) -\frac{1}{2}%
E_{J}\sigma _{x}+2E_{c}\frac{C_{J}}{e}V\sigma _{z}.  \label{ham}
\end{equation}%
where $\sigma _{x}$ and $\sigma _{z}$ are the Pauli matrices in the
pseudo-spin basis.

The time development of the state vector $|\psi (t)\rangle $ of the system
is postulated to be determined by Schr\"{o}dinger equation

\begin{equation}
i\hbar \frac{d}{dt}|\psi (t)\rangle =\widehat{H}_{in}|\psi (t)\rangle .
\end{equation}%
The solution of equation (4) can be written as $|\psi (t)\rangle =\hat{U}%
_{t}|\psi (0)\rangle ,$ where $\hat{U}_{t}$ is a time evolution operator
which depends on the Hamiltonian of the system. Since $|\psi (0)\rangle $ is
arbitrary, the time evolution operator obeys $i\hbar dU_{t}/dt=\widehat{H}%
_{in}U_{t}.$ Integrating this equation, gives $U_{t}$ $\equiv \exp \left(
\frac{-i}{\hbar }\int \hat{H}_{in}dt\right) .$ If the system is conservative
and $\hat{H}_{in}$ is explicitly independent of time, then $U_{t}$ reduces
to $U_{t}$ $\equiv \exp \left( \frac{-i}{\hbar }\hat{H}_{in}t\right) .$

One may, also, assume that the two eigenstates $|\Phi _{1,2}^{(n)}\rangle $
are known, along with their corresponding eigenenergies $\Upsilon _{\pm
}^{(n)}.$Then we can write the time evolution operator as%
\begin{equation}
\hat{U}_{t}=\sum\limits_{n=0}^{\infty }\left\{ \exp (-i\Upsilon
_{+}^{(n)}t)|\Phi _{1}^{(n)}\rangle \langle \Phi _{1}^{(n)}|\right. \left.
+\exp (-i\Upsilon _{-}^{(n)}t)|\Phi _{2}^{(n)}\rangle \langle \Phi
_{2}^{(n)}|\right\} .
\end{equation}%
In order to find explicit forms of the eigenvalues and eigenvectors, we
consider the weak quantized radiation field and then we can neglect the term
containing $V^{2}$ in equation (\ref{ham}) and using the rotating wave
approximations we obtain
\begin{eqnarray}
\Upsilon _{\pm }^{(n)} &=&\omega (n+1/2)\pm \mu _{n},  \nonumber \\
\mu _{n} &=&\sqrt{\frac{\Delta ^{2}}{4}+\frac{e^{2}C_{J}^{2}}{%
2(C_{J}+C_{g})^{2}}\times {\frac{\omega }{2\hslash C_{J}}}\times (n+1)}.
\end{eqnarray}%
We denote by $\Delta =E_{J}-\omega $ the detuning between the Josephson
energy and cavity field frequency. $|\Phi _{1,2}^{(n)}\rangle $ are given by
\begin{eqnarray}
|\Phi _{1}^{(n)}\rangle &=&\cos \left( \tan ^{-1}\left( \frac{%
(C_{J}+C_{g})\left( -\Delta +2\mu _{n}\right) }{eC_{J}\sqrt{(n+1)}}\sqrt{%
\frac{2\hslash C_{J}}{\omega }}\right) \right) |n,\uparrow \rangle  \nonumber
\\
&&+\sin \left( \tan ^{-1}\left( \frac{(C_{J}+C_{g})\left( -\Delta +2\mu
_{n}\right) }{eC_{J}\sqrt{(n+1)}}\sqrt{\frac{2\hslash C_{J}}{\omega }}%
\right) \right) |n+1,\downarrow \rangle {\Large ,} \\
|\Phi _{2}^{(n)}\rangle &=&\sin \left( \tan ^{-1}\left( \frac{%
(C_{J}+C_{g})\left( -\Delta +2\mu _{n}\right) }{eC_{J}\sqrt{(n+1)}}\sqrt{%
\frac{2\hslash C_{J}}{\omega }}\right) \right) |n,\uparrow \rangle  \nonumber
\\
&&-\cos \left( \tan ^{-1}\left( \frac{(C_{J}+C_{g})\left( -\Delta +2\mu
_{n}\right) }{eC_{J}\sqrt{(n+1)}}\sqrt{\frac{2\hslash C_{J}}{\omega }}%
\right) \right) |n+1,\downarrow \rangle {\Large .}
\end{eqnarray}%
We devote the next section to investigate the general structure of the
dipole-dipole correlation function in terms of the eigenstates and
eigenvalues of the system, and evaluate the fluorescence spectrum for input
binomial states.

\section{The transient spectrum}

For the calculation of the spectrum, we consider the time evolution of the
off-diagonal density matrix elements of the field while the diagonal density
matrix elements remain stationary. In this section we derive the physical
transient spectrum $S(\nu )$ by calculating the Fourier transform of the
time averaged dipole-dipole correlation function $\langle \sigma _{+}(t+\tau
)\sigma _{-}(t)\rangle $, weighted by the detector response function $%
\langle \psi (0)|\sigma _{+}(t+\tau )\sigma _{-}(t)|\psi (0)\rangle $ where $%
|\psi (0)\rangle $ is the initial state of the considered system. Then, the
transient spectrum is given by the expression \cite{ebe77}
\begin{equation}
S(\nu )=Re\left( \int\limits_{0}^{\infty }d\tau \exp [-i\nu \tau -\gamma
\tau ]\overline{\langle \psi (0)|\sigma _{+}(t+\tau )\sigma _{-}(t)|\psi
(0)\rangle }d\tau \right) ,
\end{equation}%
where $\gamma $ is the detector width and $\nu$ is the spectrum
frequency.

Using equations (5)-(8), the time evolution of the states can be expressed
as
\begin{eqnarray}
\hat{U}_{t}|n,e\rangle &=&A(n,t)|n,e\rangle +B(n,t)|n+k,g\rangle ,  \nonumber
\\
\hat{U}_{t}|n,g\rangle &=&A(n-k,t)|n,g\rangle +B(n-k,t)|n-k,e\rangle ,
\end{eqnarray}%
where
\begin{eqnarray}
A(n,t) &=&\sin ^{2}\left( \tan ^{-1}\left( \frac{(C_{J}+C_{g})\left( -\Delta
+2\mu _{n}\right) }{eC_{J}\sqrt{(n+1)}}\sqrt{\frac{2\hslash C_{J}}{\omega }}%
\right) \right) \exp \left( -it\Upsilon _{-}^{(n)}\right)  \nonumber \\
&&+\cos ^{2}\left( \tan ^{-1}\left( \frac{(C_{J}+C_{g})\left( -\Delta +2\mu
_{n}\right) }{eC_{J}\sqrt{(n+1)}}\sqrt{\frac{2\hslash C_{J}}{\omega }}%
\right) \right) \exp \left( -it\Upsilon _{+}^{(n)}\right) , \\
B(n,t) &=&\frac{1}{2}\sin 2\left( \tan ^{-1}\left( \frac{(C_{J}+C_{g})\left(
-\Delta +2\mu _{n}\right) }{eC_{J}\sqrt{(n+1)}}\sqrt{\frac{2\hslash C_{J}}{%
\omega }}\right) \right)  \nonumber \\
&&\times \biggl[\exp \left( -it\Upsilon _{+}^{(n)}\right) -\exp \left(
-it\Upsilon _{-}^{(n)}\right) \biggr].
\end{eqnarray}%
By using the above equations, the correlation function $\langle \hat{\sigma}%
_{+}(t+\tau )\hat{\sigma}_{-}(t)\rangle $ can be evaluated in terms of the
coefficients $A(n,t)$ and $B(n,t),$ as follows (the box initially prepared
in the excited state i.e. the initial state of the system is assumed to be $%
\left\vert \psi (0)\right\rangle =\sum\limits_{n=0}^{M}\beta _{n}^{M}(\eta
)\left\vert n,e\right\rangle ,$ where $\beta _{n}^{M}(\eta )$ will be
defined latter)%
\begin{eqnarray}
\langle \hat{\sigma}_{+}(t+\tau )\hat{\sigma}_{-}(t)\rangle
&=&\sum\limits_{n=0}^{\infty }\left( \beta _{n}^{M}\right) ^{2}\langle n,e|%
\hat{\sigma}_{+}(t+\tau )\hat{\sigma}_{-}(t)|n,e\rangle  \nonumber \\
&=&\sum\limits_{n=0}^{\infty }\left( \beta _{n}^{M}\right)
^{2}A(n,t)A(n-k,\tau )A^{\ast }(n,t-\tau ).
\end{eqnarray}%
The Fourier transform of the time averaged dipole-dipole correlation, which
is directly related to the fluorescence spectrum with the identification of $%
\gamma $ as the width associated with the detector \cite{ebe77}\textrm{.}
After carrying out the various operations we get
\begin{eqnarray}
S(\nu ) &=&\left( \beta _{0}^{M}\right) ^{2}\left( \frac{\gamma \sin
^{4}\theta _{0}}{\gamma ^{2}+(\nu -\Upsilon _{0}^{+})^{2}}+\frac{\gamma \cos
^{4}\theta _{0}}{\gamma ^{2}+(\nu -\Upsilon _{0}^{-})^{2}}\right)  \nonumber
\\
&&+\sum\limits_{n=1}^{\infty }\left( \beta _{n}^{M}\right) ^{2}{\Huge [}%
\frac{\gamma \sin ^{4}\theta _{n}\cos ^{2}\theta _{n-k}}{\gamma ^{2}+(\nu
+\Upsilon _{n-k}^{+}-\Upsilon _{n}^{+})^{2}}+\frac{\gamma \sin ^{4}\theta
_{n}\sin ^{2}\theta _{n-k}}{\gamma ^{2}+(\nu +\Upsilon _{n-k}^{-}-\Upsilon
_{n}^{+})^{2}}  \nonumber \\
&&+\frac{\gamma \cos ^{4}\theta _{n}\cos ^{2}\theta _{n-k}}{\gamma ^{2}+(\nu
+\Upsilon _{n-k}^{+}-\Upsilon _{n}^{-})^{2}}+\frac{\gamma \cos ^{4}\theta
_{n}\sin ^{2}\theta _{n-k}}{\gamma ^{2}+(\nu +\Upsilon _{n-k}^{-}-\Upsilon
_{n}^{-})^{2}}{\Huge ],}
\end{eqnarray}%
where $\left( \beta _{n}^{M}\right) ^{2}$ is the initial photon number
distribution and $\theta _{n}$ is

\[
\theta _{n}=\tan ^{-1}\left( \frac{(C_{J}+C_{g})\left( -\Delta +2\mu
_{n}\right) }{eC_{J}\sqrt{(n+1)}}\sqrt{\frac{2\hslash C_{J}}{\omega }}%
\right) .
\]%
Thus the time averaged spectrum consists of resonant structures which arise
from transitions among different dressed states. The final structure of the
time averaged spectrum will depend on the form of the input photon
distribution$.$ As the cavity field starts to interact with the Cooper pair
the initial photon number distribution starts to change. Due to the quantum
interference between component states the oscillations in the cavity field
become to be composed of two component states. The situation that has just
been described is depicted in figures 1-5.

\begin{figure}[tbph]
\begin{center}
\includegraphics[width=12cm]{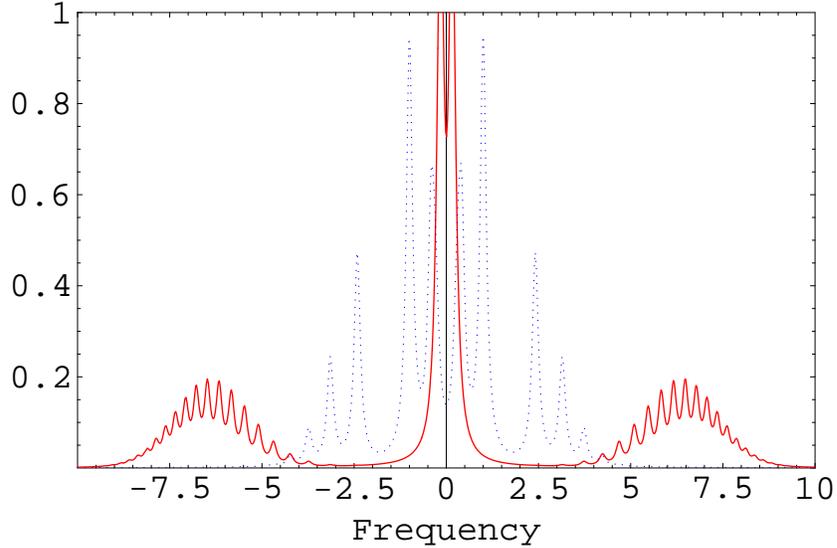}
\end{center}
\caption{The evolution of the function $S(\protect\upsilon )$ in a perfect
cavity as a function of ($\protect\upsilon -\protect\omega )/\protect\lambda
$ for an input binomial state, with the $\protect\eta =0,M\rightarrow \infty
(i.e.\protect\eta M=|\protect\alpha |^{2}$ where $|\protect\alpha |^{2}=10$
(solid line) and $|\protect\alpha |=1$( dotted line). Calculations assume
detector width $\protect\gamma =0.1g$, and the detuning parameter $\Delta $
has zero value. }
\end{figure}
In figure 1 we discuss the spectrum for the situation of a coherent field
with $\Delta =0$. We show the evolution of this spectrum $S(\upsilon )$ as a
function of ($\upsilon -\omega )/\lambda $ where $\lambda =\frac{eC_{J}}{%
2(C_{J}+C_{g})}\sqrt{\frac{\omega }{2\hslash C_{J}}}$. We consider the
initial state of the field is a binomial state. The binomial states which
has been introduced by Stoler, Saleh and Teich in [2], interpolate between
the most nonclassical states, such as number states and coherent states, and
reduce to them in two different limits. Some of their properties [2, 3, 4],
methods of generation [2, 3, 5], as well as their interaction with atoms
[6], have been investigated in the literature. The binomial state is defined
as a linear superposition of number states in an $M-$dimensional subspace
\begin{equation}
\left\vert \eta ,M\right\rangle =\sum\limits_{n=0}^{M}\beta _{n}^{M}(\eta
)\left\vert n\right\rangle ,
\end{equation}%
where the state $\left\vert n\right\rangle $ is the number state of the
field mode, $\eta $ is a complex number with the absolute value between $0$
and $1$, $M$ is a positive integer, and
\begin{equation}
\beta _{n}^{M}(\eta )=\sqrt{\left(
\begin{array}{c}
M \\
n%
\end{array}%
\right) \left( \eta ^{n}+(1-\eta )^{n-M}\right) }.
\end{equation}%
The name binomial state comes from the fact that their photon distribution
is simply a binomial distribution with probability $\eta $. The binomial
state is a linear combination of $M+1$ number states with coefficients
chosen such that the photon-counting probability distribution is binomial
with mean photon $M$. This state can produce, under certain choices of the
parameters $\eta $ and $M$, the number state $\left\vert M\right\rangle $,
the vacuum state $\left\vert 0\right\rangle $ and the coherent state. When $%
\eta \rightarrow 0,$ $M\rightarrow \infty $ \ with $\eta M=\alpha ^{2}$
fixed ($\alpha $ real constant), $\left\vert \eta ,M\right\rangle $ reduces
to the coherent states which correspond to the Poisson distribution. Here we
consider $\eta =0,M\rightarrow \infty $ and $|\alpha |^{2}=10$ (solid line)
and $|\alpha |=1$ (dotted line). Calculations assume that the detector width
$\gamma =0.1\lambda $, and the detuning parameter $\Delta $ has zero value.
The slit is adjusted so that the light is collected from Cooper-pair box
which have been in the field for times ranging from 0 up to 7. From an
initially broad featureless spectrum, the central peak and the sidebands
emerge rather quickly. As the observation region is lengthened, all the
components get taller and narrower. Then for about ($\upsilon -\omega
)\approx \pm 9\lambda $, the transient spectrum has narrowed enough and
tends to zero. In the limit of a very large mean photon number and at
exactly resonant field, it is possible to simplify the expression for the
spectrum $S(\nu )$ and illustrate explicitly the manner in which the
sidebands and the central peaks narrow with increasing ($\upsilon -\omega )$%
.
\begin{figure}[tbph]
\begin{center}
\includegraphics[width=12cm]{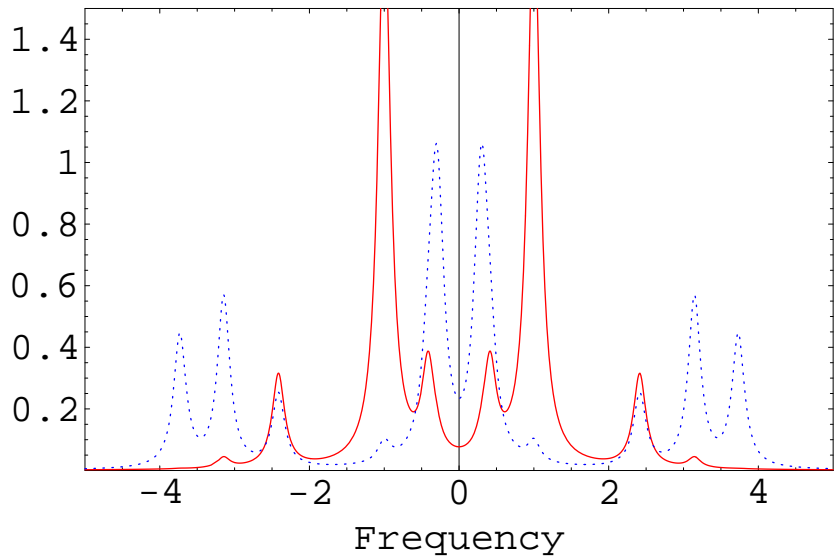}
\end{center}
\caption{The evolution of the function $S(\protect\upsilon )$ in a perfect
cavity as a function of ($\protect\upsilon -\protect\omega )/\protect\lambda
$ for an input binomial state, with $M=3$, $\protect\eta =0.7$ for (solid
line) and $\protect\eta =0.1$ for (dotted line). Calculations assume that,
the detector width $\protect\gamma =0.1g$, and the detuning parameter $%
\Delta $ has zero value. }
\end{figure}

It would be of interest to pay attention to the physical transient spectrum
for binomial states. In order to do that, we use different values of both $%
\eta $ and $M$ (say $M=3$ and $\eta =0.7,$ $0.1$). This basis is
overcomplete and many states in the Hilbert space can be expanded in it. The
behaviors of $S(\nu )$ is notably different from those observed in the
coherent state case. Indeed, in the fully connected system considered here, $%
S(\nu )$ is extremum at $\pm 1(\pm 0.4)$ when $\eta =0.7(0.1)$ whereas in
the coherent state case, figure 1, $S(\nu )$ is extremum at $\approx 0$. In
addition, the scaling behavior of the spectrum and of its derivative are
different in both cases. Therefore, when speaking about the physical
transient spectrum sensitivity achieved with different frequencies, it is
necessary to specify the kind of initial state of the field.

\begin{figure}[tbph]
\begin{center}
\includegraphics[width=12cm]{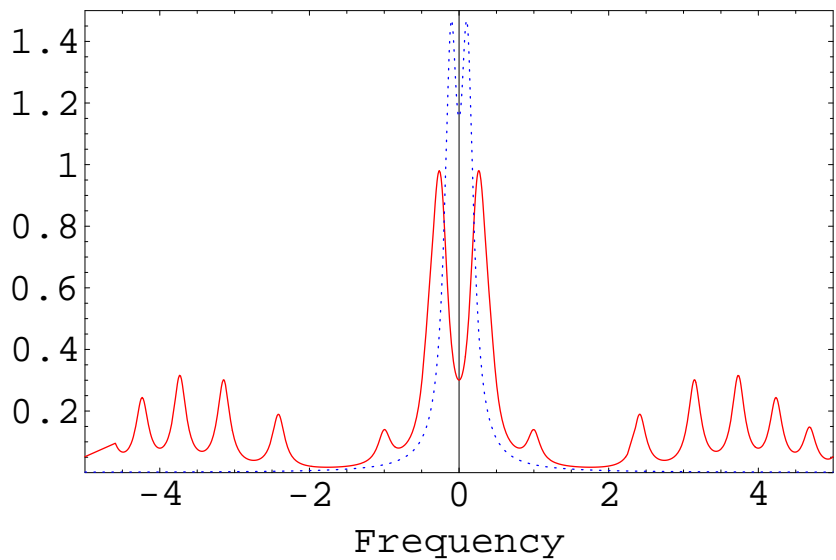}
\end{center}
\caption{The same as in figure 2 but $M=30$. }
\end{figure}

Our aim is now to evaluate the response of the physical transient spectrum
due to the mean photon $M$. \ It is interesting to note that in a large
value of $M,$ we observe a similar behavior to that obtained in the coherent
state, see figure 3. Now, we would like to shed some light on the spectrum
behavior when the detuning differs from zero. \textrm{The transient spectrum
under these conditions can be asymmetric and ultimately the central
component and one of the Rabi sidebands can vanish despite the fact that the
quadrature-noise spectrum exhibits a significant amount of noise at these
frequencies. The asymmetry arises from the stimulated emission induced
between the dressed states by the binomial field. }

\begin{figure}[tbph]
\begin{center}
\includegraphics[width=12cm]{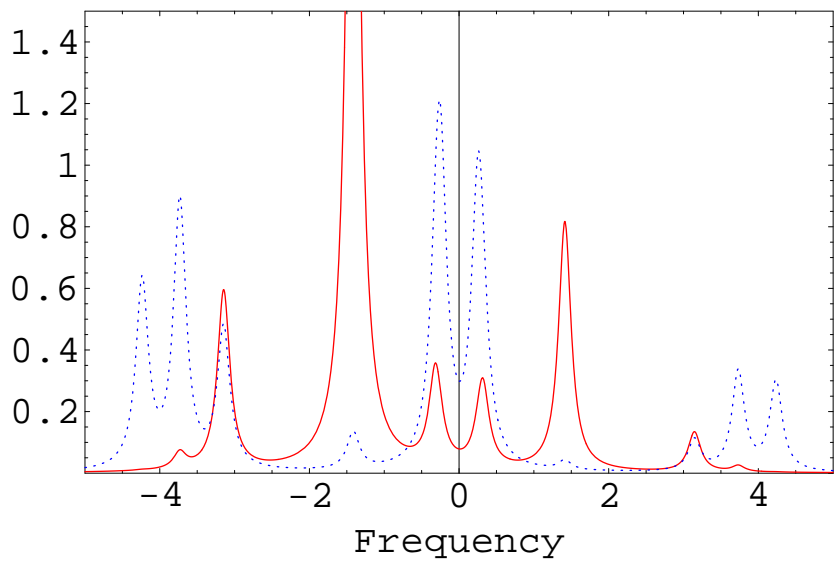}
\end{center}
\caption{The same as in figure 2 but $\Delta =g$. }
\end{figure}
These asymmetries consist of an enhanced sideband on the atomic resonance
frequency side of the central peak and more pronounced oscillations between
the central peak and the enhanced sideband than on the opposite side of the
central peak. Also, we observe a slight displacement in the location of the
central maximum from the applied field frequency toward the atomic
frequency. One can see that the large value of the detuning parameter gives
a disappearance of one peak and large displacement in the location of the
maximum value of the spectrum.

\begin{figure}[tbph]
\begin{center}
\includegraphics[width=12cm]{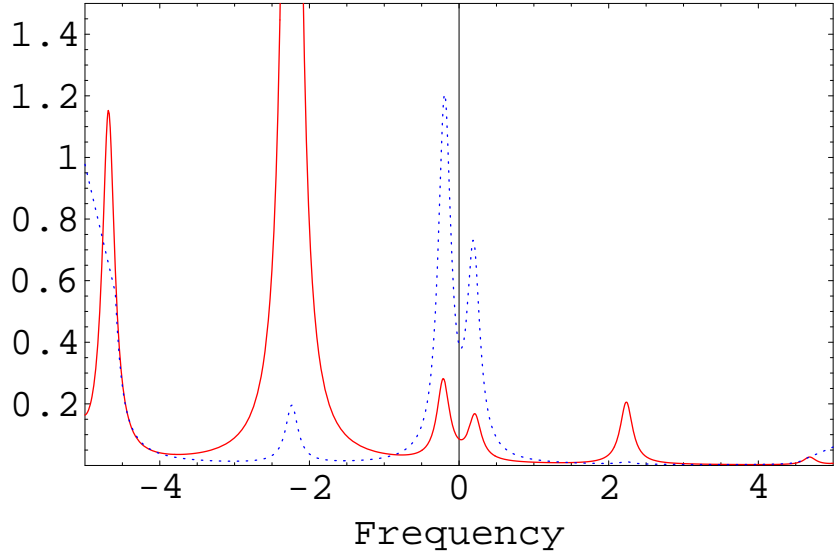}
\end{center}
\caption{The same as in figure 2 but $\Delta =2g$. }
\end{figure}
We can conclude that the effect of the detuning on the spectrum of the
emitted light is twofold. The first effect is the shift of the spectrum to
the left or to the right depending on the sign of $\Delta $. The second
effect is the dependence of the amplitude of the peaks on $\Delta $. This
dependence leads to the fact that, in the far off-resonance limit, only one
peak survives.

Finally we may point out that the present lack of a current standard based
on quantum devices has inspired several attempts to manipulate single
electrons, or Cooper pairs, where the rate of particle transfer is
controlled by an external frequency \cite{kel96}. In the experiments
described in \cite{wat03} a long array of Josephson junctions with an
external signal applied to a gate, which is capacitively coupled to the
middle of the array has been used. Theoretically many systems can act as a
qubit, but the realization of it is difficult. Nakamura et al. \cite{nak99}
have shown in their experiments with single Cooper pair box, that in a kind
of metallic island structures the oscillations between eigenstates of the
system last at least a few nanoseconds. The states are characterized by the
number of Cooper pairs in the box and (quantum) manipulation of this number
is of basic importance for the production of viable qubits. This might be a
very important issue when thinking of the limitations on the preparation and
read-out of the states.

\section{Conclusion}

We have analyzed the physical transient spectrum of a single Cooper-pair
box, which is biased by a classical voltage and irradiated by a single-mode
quantized field. We emphasize the fact that the proper expression for the
emission spectrum which is derived in this paper can be measured in a
realistic experiment. This spectrum has been obtained not only as a function
of the atomic and field parameters, but also as a function of $\gamma $
which is available to the experimenter ($\gamma $ the width associated with
the detector). Thus it correctly incorporates the possible effects of a
finite observation interval, observations made close to the point where the
interaction was turned on, and arbitrary initial conditions for the Cooper
pair box at the start. \textrm{Among the reasons for the interest in
considering nonclassical effects of the binomial states, we may mention the
following reasons: the experimental work shows that nonclassical effects
serve as a test of the quantum nature of light and nonclassical behavior of
light is usually connected with a noise reduction below a standard limit
(e.g. the shotnoise limit). }In particular, we have explored the influence
of various parameters of the system on the emission spectrum of the output
field statistics. We have used the finite double-Fourier transform of the
two-time field correlation function to find an analytical expression for the
spectrum. The spectrum in the cavity for the initial binomial states is
studied. Such systems are potentially interesting for their ability to
process information in a novel way and might find application in models of
quantum logic gates. The phenomenon of oscillations in the field spectrum
has been shown. It is observed that the symmetry shown in the resonant case
for the spectra is no longer present once the detuning is added.

\end{document}